\documentclass[12pt]{article}
\usepackage{epsfig,psfig,graphicx,rotating}
\newcommand{\be}{\begin{equation}}
\newcommand{\ee}{\end{equation}}
\newcommand{\bea}{\begin{eqnarray}}
\newcommand{\eea}{\end{eqnarray}}

\begin{document}
\begin{titlepage}

\centerline {\Huge {\bf Extreme Energy Cosmic Rays:}}

\bigskip

\centerline {\Huge {\bf Bottom-up vs. Top-down
    scenarii\footnote{\large{Based on lectures at the 
Fourth International Workshop on `New Worlds in Astroparticle Physics'
in Faro, Portugal, September 2002,  at the 9th Course on
    Astrofundamental Physics of the Chalonge School, Palermo, Italia,
    September 2002 and at the SOWG EUSO meeting,
    Roma, Italia, November 2002. } }} }



\begin{center}
 { \Large {\bf H. J. de Vega$^{(a),} $\footnote{devega@lpthe.jussieu.fr},  
N. Sanchez$^{(b),}\footnote{Norma.Sanchez@obspm.fr}$}} \\ \vskip 15.0pt
(a)LPTHE, UMR 7589, CNRS,
Universit\'e Pierre et Marie Curie (Paris VI) \\
et Denis Diderot  (Paris VII),  Tour 16, 1er. \'etage, \\ 4, Place Jussieu
75252 Paris, Cedex 05, France \\
(b) Observatoire de Paris,  LERMA, UMR 8540, CNRS\\ 61, Avenue de
l'Observatoire, 
75014 Paris,  France. \\
\end{center}

\vskip 30.0pt

\centerline {\bf Abstract}

\noindent
We present an overview on extreme energy cosmic rays (EECR) and the 
fundamental physics connected with them. The top-down and bottom-up scenarii 
are contrasted. We summarize the essential features underlying the top-down 
scenarii for EECR, namely, the lifetime and the mass {\bf imposed} to
the heavy relics 
whatever they be: topological and non-topological solitons, X-particles, cosmic
defects, microscopic black-holes, fundamental strings. An unified formula for 
the quantum decay rate of all these objects was provided in hep-ph/0202249. The
key point in the top-down scenarii is the necessity to {\bf adjust}
the lifetime of the heavy object to the age of the universe. The
natural lifetimes of such heavy objects are, however, 
microscopic times associated to the GUT energy scale ($\sim
10^{-28}$sec. or shorter); such heavy objects could have been
abundantly formed by the end of inflation and it seems natural they
decayed shortly after being formed.  The arguments produced to  {\bf fine
tune} the relics lifetime to the age of the universe are critically
analyzed. The annihilation scenario (`Wimpzillas') is analyzed too.
Top-down scenarii based on networks of topological defects are strongly 
disfavored at the light of the recent CMB anisotropy observations. 
We discuss the acceleration mechanisms of cosmic rays,
their possible astrophysical sources and the main open physical problems and
difficulties in the context of bottom-up scenarii, and we conclude by
outlining the expectations from future observatories like EUSO and
where the theoretical effort should be placed. 
\vskip 3cm

\end{titlepage}

\section{Introduction}

Cosmic rays are one of the rare systems in the Universe which are not
thermalized. Their energy spectrum follows approximately a power law
over at least thirteen  orders of magnitude.
The understanding of the cosmic ray spectrum involves several branches
of physics and astronomy.

First, to explain how cosmic rays get energies up to $10^{20}$eV
according to observations and with such power
spectrum. Then, to study the effect of galactic and extragalactic
magnetic fields and explain the knee and the ankle effects. 
Furthermore, clarify whether the GZK effect is present or not in the
cosmic rays events observed beyond $10^{20}$eV. Last and not least to
understand the interaction of the cosmic rays with the atmosphere, the
extended air shower formation, especially at extreme energies and the
fluorescence effects. At the same time, one has to identify the
astronomical sources of cosmic rays and relate the observed events,
composition and spectra with the properties and structure of the sources.

The standard physical acceleration mechanism goes back to the ideas
proposed by E. Fermi in the fifties. That is, charged
particles can be efficiently accelerated by electric fields in
astrophysical shock waves\cite{raycos,tom}. This is the so-called
diffusive shock acceleration mechanism yielding a power spectrum with 
\be \label{esp}
n(E) \sim E^{-\alpha} \; ,
\ee
with $ 2.3 \leq \alpha\leq 2.5-2.7 $. Such spectrum is
well verified over thirteen orders of magnitude in energy. 

\begin{figure}[ht]
\centerline{\epsfxsize=14.0cm \epsfbox{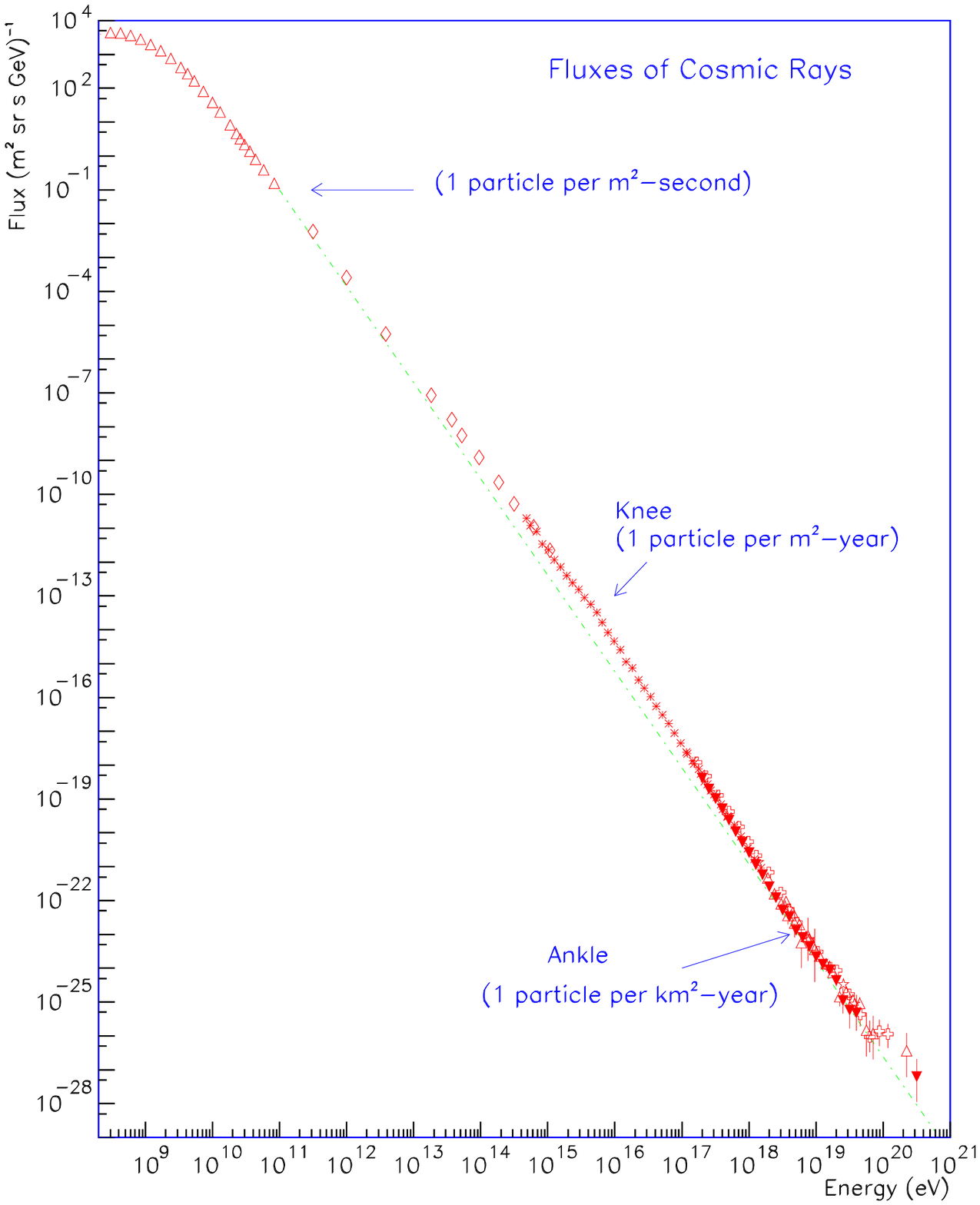}}  
\caption{ Cosmic ray spectrum[22]. }
\label{fig1}
\end{figure}

\section{Top-down scenarii}

Top-down scenarii for extreme energy cosmic rays (EECR) are based on
heavy relics from the early 
universe which are {\bf assumed} to decay at the present time or from
topological defects also originated in the early universe. For all relics,
(whatever their nature: heavy particles, topological and
non-topological solitons, 
black-holes, microscopic fundamental strings, cosmic defects etc.), one
has to {\bf fine tune} the lifetime of these objects to be the age of the 
universe. 

It has been further proposed that stable keavy relics can produce EECR
through annihilation by pairs\cite{chung,rocky}. This scenario suffer
a different type of inconsistencies as we show below.

The second type of top-down scenarii  rely on the existence of a network of
topological defects formed during phase transitions in the early
universe. Such topological defects should survive till nowadays to
produce the observed EECR. In case they decay in the early universe we
go back to the previous case. It must be first noted that only some
grand unified field theories support topological
defects\cite{nos}. Moreover, recent CMB anisotropy measurements from
Boomerang, Maxima, Dasi and Archeops\cite{CMB}  have seen no
evidence of topological defects strongly disfavoring their eventual
presence in the present universe.

We provided in ref.\cite{nos} an unified description for the quantum
decay formula of unstable particles which encompass all cases
above mentioned as well as the particle decays in the standard model
(muons, Higgs, etc). In all cases the decay rate can be written as,
\be \label{formae}
\Gamma = \frac{g^2 \; m }{\mbox{numerical factor}}
\ee
where $g$ is the coupling constant, $m$ is the typical mass in the theory
(it could be the mass of the unstable particle) and the numerical
factor contains often relevant mass ratios for the decay process.

The {\bf key drawback} of all top-down scenarii is the {\bf lifetime problem}.
The {\bf ad-hoc} requirement of a lifetime of the order the age of the universe
for the heavy particles implies an operator with a very high dimension
describing the decay, and/or an extremely small coupling constant (see below).

Heavy relics could have been formed by the end of inflation at typical
GUT's energy scales, but their {\bf natural} lifetime would be of the order
of {\bf microscopic} times typically associated to GUT's energy scales
\cite{big}. 

The problem in the top-down scenarii is {\bf not} the formation
of heavy particles or topological defects. They could all have been generated 
in the early universe. The key problem is their existence {\bf today} (i. e.
their imposed lifetime of the order of the age of the universe) and
the value of their mass that must be adjusted to be $ \sim 10^{20}$ eV.

\bigskip

Heavy particles with masses in the GUT scale can be produced in large
numbers during inflation and just after inflation\cite{big}. The production
mechanism is particle production by the expanding metric and
parametric or spinodal amplification in the inflaton field. That is, 
linear resonance of the quantum modes of the heavy field in the
background or condensate of the inflaton. In addition, non-linear quantum
phenomena play a crucial role and can enhance the particle
production\cite{big}. Such non-linear production is of the same order
of magnitude as the gravitational production of particles by the time
dependent metric. 

\bigskip

Once these heavy particles are produced they must have a lifetime of
the order of the age of the universe in order to survive in the
present universe and decay into EECR. Only in the early
universe the production of such heavy objects is feasible due to their
large mass.  

Moreover, in order to be the source of  EECR, these
particles must have a mass of the observed EECR, namely $ m_X > 
10^{20}$ eV = $ 10^{11}$ GeV. 

\bigskip

The effective Lagrangian describing the
X-relic decay  contains a local monomial of dimension $n$ (in mass units) 
\be \label{lefX}
{\cal L}_I = {g \over M^{n-4}} \, X \Theta \; ,
\ee
here the field $ X $ is associated to the decaying particle of mass $
m_X $ and $ \Theta $ stands for the product of fields coupled to it.

Then, the  decay rate for a particle of mass $ m_X $ takes the form\cite{nos}
\be \label{anchX}
\Gamma = { g^2 \over \mbox{numerical factor}}  \; m_X \; \left( {m_X
\over M}\right)^{|2n-8|}
\ee
and  their lifetime will be given by 
$$
\tau_X = { \mbox{numerical factor}\over g^2 } \; { 1 \over m_X} \;
\left( {M \over m_X}\right)^{2n-8} =  { \mbox{numerical factor}\over
g^2 } \; { 1 \over m_X} \; 10^{6(n-4)} 
$$
where we set a GUT mass $ M = 10^{15}$ GeV. The age of the
universe is $ \tau_{universe} \sim 2 \; 10^{10}$years and we have to
require that $ \tau_X > 
\tau_{universe} $. Therefore,
\be \label{taunive}
10^{54} < 
{ \mbox{numerical factor}\over g^2 } \; 10^{6(n-4)}
\quad \mbox{or}  \quad  \log_{10} g < 
3 (n - 13)
\ee
where we dropped the numerical factor in the last step.

For $ g \sim 1 $, eq.(\ref{taunive}) requires an operator $
\Theta $ with dimension at least thirteen in the effective Lagrangian
(\ref{lefX}) which is a pretty high dimension. 

That is, one needs to exclude all operators of dimension lower than
thirteen in order to 
extremely suppress the decay. Clearly, one may accept lower dimension operators
$ \Theta $ paying the price of a small coupling $ g $. For example: $
g = 10^{-9} $ and $ n = 10 $ fulfill the above bound still being a
pretty high dimension operator.  Notice that a moderate $n$ as $n=4$
lowers the coupling to $g \sim 10^{-27} $.

In summary, a heavy X-particle can {\bf survive} from the early
universe till the present times {\bf if} one chooses

\begin{itemize}
\item{an extremely small coupling $g$}

$\quad  $             and/or

\item{an operator $\Theta$ with high enough dimension}
\end{itemize}

{\bf None} of these assumptions can be supported by arguments other than
imposing a lifetime of the age of the universe to the X-particle. That
is, the lifetime must be here {\bf fine tunned}. That is, one has to
built an {\bf ad-hoc} Lagrangian to describe the X-particle decay. 
Indeed, a variety of  {\bf ad-hoc} lagrangians have been proposed in the
literature together with the symmetries which can {\bf adjust} a wide
variety of lifetimes\cite{discrete}. 

\bigskip

In order to cope with  the lifetime problem a 
number of so called `solutions' have been invoked, but all the  `remedies' 
replace an assumption  by another one, namely:
\begin{itemize}
\item{(1)}  an assumed new global symmetry to protect the X particle and 
which would be only broken non-perturbatively by quantum gravity 
wormholes \cite{wormholes} or instanton-type  effects \cite{instanton}
to make $\tau_X=$ age of the universe.
However, these quantum gravity effects are poorly controlled, basic
theoretical  
uncertainties remain, (the sign of the euclidean gravity action being 
only one of them), which would produce the opposite effect to the one 
claimed; thus $\tau_X$ would be exponentially shortened (instead of
increased) by wormholes.

\item{(2)} discrete gauge symmetries to construct high dimensional operators 
\cite{discrete}. As stated before, these are all ad-hoc lagrangians
built on the  
assumption that such group symmetries could have a physical r\^ole. No
fundamental physical reason  exists to argue for them.
\item{(3)} On the same line of thinking, fractionally charged particles 
(`cryptons') have been invoked \cite{cryptons} from some particularly chosen 
hidden sectors of particular effective string/M inspired models. 
Then, support to the assumption of a long lifetime for $\tau_X$ would 
come from a strongly interacting (bound state) sector
and its non-pertubative dynamics (which is not controlled), in 
flipped $SU(5)$ for instance. But, as many `particle' sectors 
appearing in string inspired phenomenology, no physical reason exists 
to choose for such states, in particular for fractionally charged 
(bound) state particles.
\end{itemize}
Finally, in the line of reasoning of (1)-(3) mentioned before, comparison 
with the stability or `metastability' of the proton  has been invoked 
in order to support for the stability (and decay) of the X-particles. 
In the standard model, in which  baryon-lepton number is 
conserved, proton decay can be realized only by ad-hoc introducing 
non-renormalizable high dimensional operators; then, it is invoked 
that `a new global symmetry' for the X-particle could exist, only 
broken by operators suppressed by $M^n$ with $M= M_{planck}$ and $n>7$
\cite{kach}. 

In  other words, ad-hoc proton decay  is argued to support ad hoc 
X-particle decay.
As is known, GUT models predict proton decay (which have not been 
found so far), placing a lower bound of the proton lifetime 
$\tau_{proton} > 1.6 \, 10^{25}$yr. Proton decay is however a natural
consequence of grand unification as lepton and quarks belong to the
same multiplet. 

\bigskip

A common feature to top-down approach is that the arguments trying 
to support a long  lifetime
for the X-particles successively call for  more and more speculative
explanations.

\bigskip

Still, the essential question in the top-down scenarii remain, i.e.: IF 
X-particles and topological defects
in such scenarii have NOT decayed in the early universe, shortly 
after they formed, WHY they
should decay JUST now?, i.e. the lifetime {\bf fine tuning} remains.

\bigskip

The top-down scenarios  are just tailored to explain the observed 
events. There is absolutely no physical reason to assume that relics 
have such a mass (and not any other value) and such a lifetime.

\bigskip

Moreover, the  question of stability of topological solutions is a highly non 
trivial issue. The mere presence of a conserved topological charge 
does {\bf NOT} guarantee their stability, the energy must be related to the 
topological charge and must be bounded from below by the topological 
charge. Otherwise, the topological defect is  {\bf unstable}.

\bigskip

Closed vortices from abelian and non-abelian gauge theories are {\bf not}
topologically stable in $3+1$ space-time dimensions. Static vortices
in $3+1$ space-time dimensions just collapse to a point since their
energy is proportional to their length. They do that in a very short
(microscopic) time. 

It must be noticed that only a restricted set of spontaneously
broken non-abelian gauge theories exhibit vortex solutions. For
example, there are no topologically stable vortices in the standard
$SU(3) \times SU(2)  \times U(1) $ model in $3+1$ space-time
dimensions. Grand unified theories may or may not 
posses vortex solutions in $2+1$ space-time dimensions
depending under which  representations of the
gauge group belong the Higgs fields.

Cosmic strings are closed vortices of horizon size. 
In $3+1$ space-time dimensions, strings collapse very fast
except if they have horizon size in which case their lifetime would be
of the order of the age of the universe. However, such horizon size
cosmic strings are excluded by the CMB anisotropy observations and by
the isotropy of cosmic rays. 

Such gigantic objects would behave classically whereas microscopic closed
strings (for energies $ < 
M_{Planck} = 10^{19}$ GeV) behave quantum mechanically. 

The existence of cosmic string networks is not established although
they have been the subject of many works. In case such networks would
have existed in the early universe they may have produced heavy
particles X of the type discussed before and all the discussion on
their lifetime applies here. The discussion on the lifetime problem
also applies to rotating superconducting strings which  have been proposed as
classically stable objects\cite{vort}. 

In summary, a key point here is the {\bf unstability} of topological defects in
$3+1$ space-time dimensions. Unless one chooses very specific models
\cite{topnotop}, topological defects decay {\bf even classically} with a
short lifetime. They collapse to a point at a speed of the order of
the speed of light. 

\subsection{Annihilation Top-down Scenarii}

There are top-down scenarii where, instead of decay, annihilation of
the relic superheavy particles (wimpzillas) have been
proposed\cite{chung,rocky}. 
That is, in this scenario the relics with mass $M_X \sim 10^{12}$GeV
are stable and produce EECR through annihilation when they
collide\cite{chung,rocky}. Here, the lifetime free parameter is
replaced by the annihilation cross section. 
These superheavy particles are assumed to be produced during
reheating\cite{chung}. Its annihilation cross section is thus bounded
by the amount of dark matter in the universe: $ \sigma_X \sim \alpha
\; (M_X)^{-2} $ and $ \alpha \leq 0.01 $. In 
ref.\cite{rocky} the produced EECR flux is computed  for several
scenarii:

For a smooth dark matter distribution assuming a NFW singular profile
it is required that $ \sigma_X = 6 \times 10^{-27}$cm${}^2$ in order to
reproduce the observed EECR flux\cite{rocky}. But this value for  $
\sigma_X $ is
$10^{27}$ larger than the maximun value compatible with $\Omega \sim 1
$ in the early universe (corresponding above to $ \alpha \sim 0.01$).

In a second scenario, dark matter is assumed
to form into clumps with NFW profiles. In this way, the EECR flux
is about $\sim 2000 $ times larger due to the increase of the dark matter
density\cite{rocky}. However, that needs again a  value for $ \sigma_X \; \;
  10^{24}$ times larger than its value in the early universe. 

Finally, isothermal clumps have been considered. There, the  EECR flux
turns to be independent  of $ \sigma_X $ but the 
internal radius of the clump $R_{min}$ is proportional to $ \sigma_X $. 
For our galaxy, $R_{min}$ turns to be $\sim 10^{-42}$ times the size
of the clump. For a $10$kpc clump this gives $R_{min}\sim
10^{-20}$cm which is a high energy quantum microscopic scale. 
A further problem in this scenario raised in ref.\cite{rocky}
is that the predicted EECR flux is too high by a factor $10^{15}$.
Then, in order to reduce the flux, it is proposed
in ref.\cite{rocky} that these wimpzillas amount only to $10^{-15}$
of the dark matter. This can be achieved by setting $ \alpha \sim
10^{-17}$ in the annihilation cross section. But such  $ \sigma_X $
makes a problem for $R_{min}$ which becomes then too small $ \sim
10^{-37}$cm $\sim 10^{-4}$Planck length. 

\subsection{Signatures of top-down scenarii}

We have discussed above the main points from which  the top-down approach 
can be theoretically criticized.
Let us now mention some characteristic features of top-down models 
which can be taken as signatures to constraint or disclaim them from
observational data: 

\begin{itemize}

\item{(1)} Spectra of the particles generated in the top-down models are 
typically flatter than the bottom-up ones. Contrary to the
acceleration mechanisms, the top-down generated spectra do not follow
a power law. 

\item{(2)} The composition of the EECR's at the source in the top-down 
scenarii is  dominated by gamma rays and 
neutrinos (only $5\%$ of the energy is in protons). Although propagation 
over cosmological distances modify the ratio of gamma rays to protons,
still photons considerably dominate over protons.
Top-down scenarii could be constrained by the cascade produced at low 
energies (MeV-GeV) by the gamma rays originated at distances larger 
than the absorption length \cite{bhasi}.

\item{(3)} Fluxes of EECR's provided by topological defect models are much 
lower than required. Simulations of cosmic string networks including 
self-intersection,  inter-commutation, multiple loop fragmentation, 
as well as cusp annihilation, all produce too low fluxes as compared 
with observations \cite{bhasi,berez}.  (Some simulations of long strings 
networks 
claiming flux enhancement were recently discussed, but the 
typical distance between two such string segments being the Hubble 
scale, the EECR's produced in this way would be completely 
absorbed\cite{bhasi,berez}. 
And, if by chance, a string as such would be near us (about few tens 
Mpc), it would imply a large anisotropy in EECR events, which is is 
not observed).
In any case, simulations of the dynamics of cosmic string networks 
(with the cosmic expansion included)  are not well controlled, making 
them not predictive.
\end{itemize}
Let us recall that the recent  CMB anisotropy observations \cite{CMB}
strongly disfavor topological defects in the present universe.(We have
discussed here topological defects since they are still 
considered  in the EECR's  top-down literature).

\section{Bottom-up scenarii}

EECR may result from the acceleration of protons and ions by
shock-waves in astrophysical plasmas (Fermi acceleration
mechanism)\cite{raycos}. Large enough sources or small sources with
strong magnetic fields can accelerate particles to the
energies of the observed EECR. Sources in the vicinity of our galaxy
as hot spots of radio galaxies (working surfaces of jets and the inter
galactic medium) and blazars (active galactic nuclei with relativistic
jet directed along the line of sight) as BL Lacertae can evade the GZK
bound\cite{floyd,TT,raycos}. 

\bigskip

The acceleration of charged particles like protons and ions take place in 
astrophysical shock waves.
In short, electric fields accelerate charged particles in the
wavefront of the shock. Then, magnetic fields deviate and diffuse the
particles. Charged particles {\bf take energy} from the wavefront of
macroscopic (astronomic!) size. Particles trapped for long enough time
can acquire gigantic energies. This mechanism can accelerate particles
till arbitrary high energies. The upper limit in energy is given by
the time during which the particle stays in the wavefront which depends on
the size of the source and on the magnetic field strength. 

\bigskip

Particles gain energy by bouncing off hydromagnetic disturbances near
the shock wave. Particles are both in the downstream and the upstream
flow regions. Assume that  a particle in the upstream crosses the shock-front,
reaches the downstream region and then {\bf crosses back} to the
upstream region. This double crossing by the particle boosts its energy 
by a factor proportional to the square of the Lorentz factor of the 
shock $ \sim \Gamma^2 $ \cite{raycos}. Such factor may be very large, 
indeed, the larger is the acceleration energy, the less probable is the 
process, finally yielding a power spectrum that decreases with the energy 
[see eq.(\ref{esp})].

\bigskip

Accelerated particles are focalized inside a cone of angle $ \theta \leq
\frac{1}{\Gamma} $. To rotate the particle momentum $ \vec p $ this
angle takes a time $ \Delta t \sim r_g \;  \theta = \frac{E}{Z
  \, e \, B_{\perp} \, \Gamma} $ where $ r_g =\frac{E}{Z \, e \,
  B_{\perp}} $ stand for the relativistic gyration radius of the
particle in the magnetic field. It must be $ \Delta t < R_s $ where
$ R_s $ is the radius of the spherical wave, otherwise the particle is
gone of the shock front. Therefore, $ E < Z  \, e \, B_{\perp} \,
\Gamma \,  R_s $. That is, one finds that the maximal available acceleration
energy\cite{raycos,tom} for a particle of charge $Z \, e $ is,
\be\label{cota}
E_{\mbox{max}} =  Z  \, e \, B_{\perp} \,\Gamma \, R_s
\ee
There are other estimates but all have the same structure. The maximal
energy is proportional to the the magnetic field strength, to the
particle charge, to $ R_s $ which is of the order of the size of
the source and to a big numerical factor as $\Gamma$. 

\bigskip

Further important effects are the radiation losses from the
accelerated charged particles and the back-reaction of the particles
on the plasma. That is, the non-linear effects on the shock wave. 

\bigskip

Particle acceleration in shock-waves can be described at different
levels. The simplest one is the test particle description where the
propagation of charged particles in shock-waves is studied. A better
description is obtained with transport equations. In addition,
non-linear effects can be introduced in such a Fokker-Planck
treatment\cite{raycos,tom}. 

The distribution function for the particles in the plasma $ f({\vec
  x}, {\vec p}, t) $ obeys the Fokker-Planck equation,
\be \label{fopa}
\frac{\partial f}{\partial t} + {\vec u}\cdot {\vec\nabla} f -
\frac{p}{3} \; \frac{\partial f}{\partial p} \; \mbox{div} \; {\vec u} -
\mbox{div}\left( \kappa \; \nabla f\right) = Q  \; .
\ee
Here, $ \vec u $ stands for the velocity field, the third term
describes the adiabatic compression and it follows from the collision
terms in the transport equation for small momentum transfer, $  \kappa
$ describes the spatial diffusion and $ Q $ is a injection or source
term. 
The energy spectrum follows from this equation irrespective of the
details of the diffusion. It must be recalled that the coefficients in
this Fokker-Planck equation are only sketchily known for relevant
astrophysical plasmas. A microscopic derivation of the Fokker-Planck
equation including reliable computation of its coefficients will be 
important to understand the acceleration of extreme energy cosmic rays
(EECR).

Let us consider stationary solutions of the Fokker-Planck equation
(\ref{fopa}) for a simple one dimensional geometry. Let us consider a
step function as velocity field\cite{raycos}. That is,
\be \label{campov}
v(x) = v_1 \quad \mbox{upstream}, \; \mbox{for} \; x<0
\quad\mbox{and}  \quad v(x) = v_2 \quad \mbox{downstream}, \;
\mbox{for} \; x>0 \; . 
\ee
Eq.(\ref{fopa}) then takes the form,
$$
v(x) \; \frac{\partial f}{\partial x}  = \frac{\partial }{\partial
  x}\left[ \kappa(x,p) \frac{\partial f}{\partial x} \right] \quad
\mbox{for} \;  x \neq 0 \; .
$$
Integrating upon $x$ taking into account eq.(\ref{campov}) yields,
$$
v(x) \; f(x,p) =  \kappa(x,p) \; \frac{\partial f}{\partial x} + A(p)
\; ,
$$
where $A(p)$ is an integration constant. Integrating again upon $x$
gives the solution
\be \label{sigts} 
f(x,p) =
\left\{
\begin{array}{l}
f_1(p) + g_1(p) \; e^{v_1\int_0^x\frac{dx'}{ \kappa(x',p)}} \; ,
\mbox{upstream}, \; x<0 \\ \\
f_2(p) \; , \; \mbox{downstream},  \; x>0 \; . 
\end{array}  \right.
\end{equation}
Matching the solutions at the shock-wave front at $ x = 0 $ yields,
\be \label{eqp}
(r-1) p \; \frac{\partial f_2}{\partial p} = 3 \; r \; (f_1-f_2) \quad
, \quad g_1 = f_2 - f_1 \quad  \mbox{and} \; r \equiv \frac{v_1}{v_2}
\; .
\ee
Notice that the solution is independent of the diffusion coefficient $
\kappa(x,p) $. Eq.(\ref{eqp}) has the homogeneous solution (no
incoming particles),
$$
f_2(p) = A \; p^{-a} \quad ,  \quad a \equiv \frac{3 \; r}{r-1} \; .
$$
For ultrarelativistic shock-waves we have $ r = 3 $ and $ a = 9/2
$\cite{ldl}. Therefore, the cosmic ray (CR) energy flux follows the law
$$
n(E) = p^2 \; f_2(p) = A \; p^{-\frac52} \; ,
$$
in close agreement with the observations.

\bigskip

It must be stressed that this example is quite oversimplified. The
geometry is not one-dimensional in astrophysical plasmas, the velocity
field is not uniform, neither constant. However, the results obtained
show the robustness of the approach\cite{raycos,tom}.

\bigskip

The particles back-reaction on the shock-wave can be neglected
provided: (a) the shock thickness is much smaller than the CR mean free
path and (b) the CR energy is much smaller than the shock
energy. Otherwise, one has to take into account the back-reaction by
coupling the transport equation for the CR with the hydrodynamic
equations for the plasma including the CR pressure. The parameter that
measures this non-linear effect is the so called injection parameter:
$ \nu \equiv n_{CR}/N_1(upstream) $. For $ \nu \leq \nu_c \sim 0.001 $
the linear theory can be used. For  $ \nu > \nu_c $ intrinsically
nonlinear phenomena can show up as solitons and perhaps self-organized
criticality. The non-linear effects predict a maximum CR energy and a
hardening of the energy spectrum near this maximum\cite{raycos}.

\subsection{Astrophysical Sources}

Let us now discuss the possible astrophysical sources of EECR, both
charged particles (protons, ions) and neutrinos. 
\begin{itemize}
\item{(1)}
According to eq.(\ref{cota}) sources of large size and/or large
magnetic fields are needed. For big sources natural candidates are
active galactic nuclei (AGN). These are supermassive black holes (mass
$ \sim 10^6 - 10^9 $ solar masses) in the core of quasars. They are
powered by the matter accreting onto the core black hole.  It has been
found from X and $\gamma$ ray detection that they are Kerr  black
holes near their extreme limit, that is with $J = G \; M^2/c$. 
[These black-holes should have formed by rotating collapsing matter. 
Their angular momentum is large because it
  is  just conserved during collapse; the excess should be radiated by
  gravitational radiation and what remains is an extreme or near extreme
 Kerr black-hole]. They usually exhibit powerful jets identified through 
radio emission where shock-waves can accelerate CR's.  
For example, acceleration may happen in radio lobes of radio 
galaxies\cite{raycos,halzen}. 
The interaction regions may extend as much as a Mpc. The jet energy is
dissipated into a bow shock inside the inter galactic medium and in
shocks within the jet plasma itself. The characteristic size of the
shock is about $ R_s \sim 10$kpc and the magnetic field is 
$ B \sim 10-100\mu$G which yields $E_{\mbox{max}}\sim 10^{21}$eV \cite{tom}.

Shock-waves in the inner edge of accretion disks around supermassive
black-holes in AGN can accelerate protons and ions to EECR
energies. These particles interacting with the dense photon field
yields extreme energy neutrinos. While neutrinos are emitted and can
reach us, the protons and ions are absorbed there.

\item{(2)}
Blazars (quasars, with their jets pointing toward us) are good
candidates like BL Lacertae. They have lower matter density that
facilitates propagation of the CR.

There are a few known radio jets inside the GZK sphere. We have
Centaurus A at about 3 Mpc away and M87 at 18 Mpc from us. Other
highly luminous sources as Cygnus A are too far $ \sim 200$Mpc. In
order to explain the apparently isotropic distribution of detected
events one needs an abundant and uniform distribution of sources or
intergalactic magnetic fields able to isotropize the CR flux. 

Radiou loud quasars are strong gamma emitters. If the gammas are
produced by relativistic nuclei, high energy neutrinos can be
produced too. 

\item{(3)}
Extragalactic neutron stars, in particular magnetars (fastly rotating
young neutron stars with $ B \sim 10^{15}$Gauss and $\Omega  \sim 10^4
\, s^{-1}$)  may be small size sources of EECR. Galactic neutron stars
would give a CR spectrum concentrated in the galactic plane unless strong
galactic magnetic fields could isotropize the CR flux\cite{tom,estneu}.
\item{(4)}
Sources of gamma ray bursts (GRB) may also accelerate CR since they have
strong ultrarelativistic shock-waves\cite{grb}. However, 
the ability of GRB sources to provide enough energy input to account
for EECR  observed beyond $3 \, . \, 10^{19}$eV (from AGASA) is criticized in
\cite{scst}. (See in addition \cite{wax}). 

\end{itemize}

\begin{figure}[ht]
\centerline{\epsfxsize=14.0cm \epsfbox{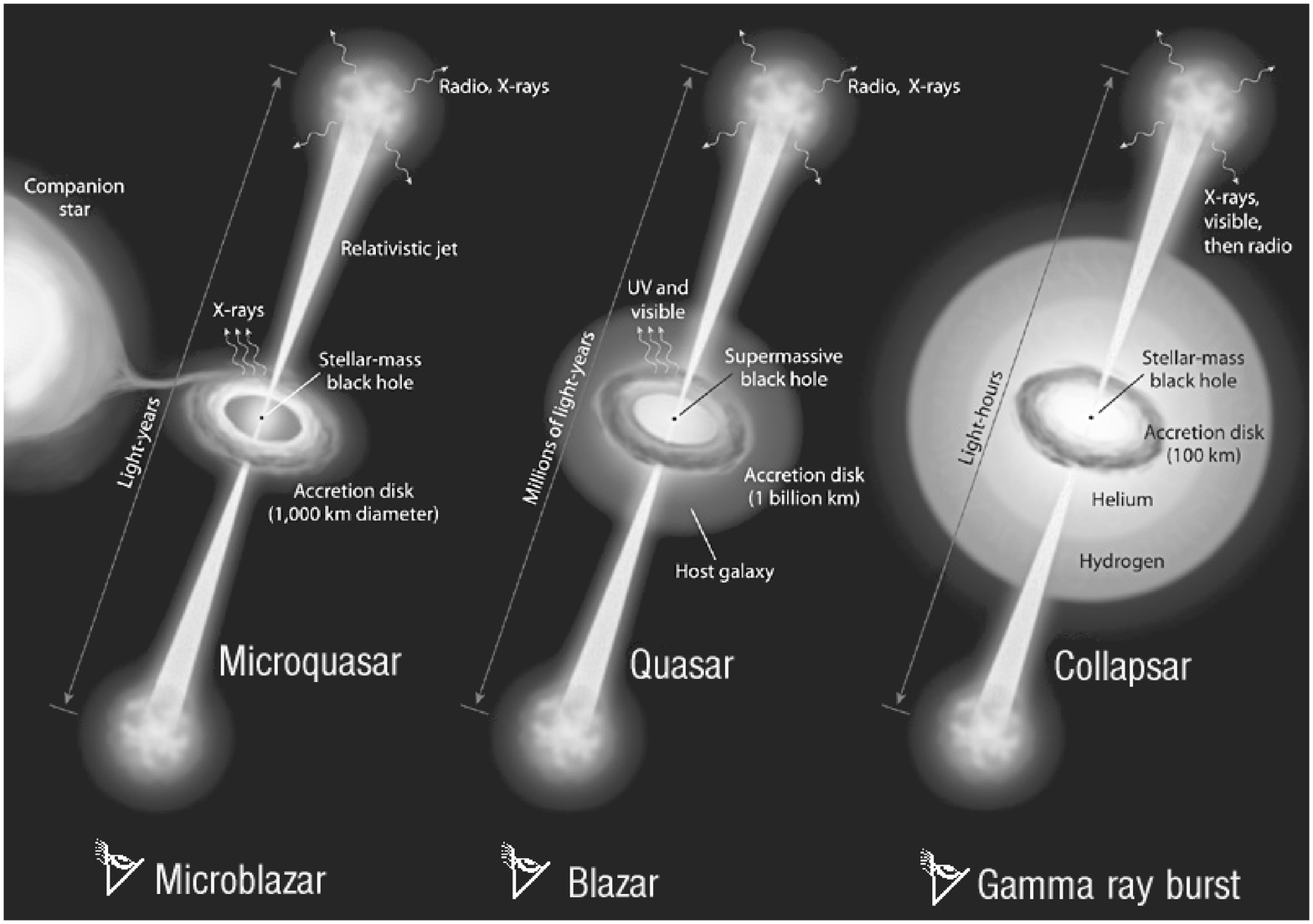}}  
\caption{Diagram illustrating current ideas concerning microquasars,
  quasars and gamma-ray burst sources (not to scale) extracted from
  ref.[14]. It is proposed in ref.[14] (see also ref.[15]) that a
  universal mechanism may be at work in all sources of relativistic
  jets in the universe. }
\label{fig2}
\end{figure}

The present data from HiRes and AGASA seem incompatible above
$10^{20}$eV where  AGASA has eight events beyond the GZK
bound. Except for these AGASA events, the available EECR spectrum today
seems compatible with the  GZK effect showing up as
predicted\cite{floyd,raycos}. 

\section{What should we expect from the forthcoming observatories like
  EUSO?}

It is 
the task of the forthcoming experiments like Auger and EUSO to clarify
the situation and show whether the  GZK effect is present or not.

\bigskip

Furthermore, Auger and EUSO should be able to see or reject the pileup
effect. The pileup effect is an enhancement of the CR spectrum just
below the GZK cutoff due to CR starting out at higher energies and
crowding up at or below the GZK energy\cite{floyd}. 

\bigskip

In addition, the measurement of the CR spectrum at GZK energies will
provide an unprecedent strong check of Lorentz invariance (or the
discovery of its violation) since the Lorentz factor involved
is extremely large\cite{floyd} $ \sim 10^{11}$.

\bigskip

Last but not least, future observations should clarify the composition
of the EECR. Neutrinos should be detected if present in the EECR. We
have seen that there are potential astrophysical sources of extreme
energy neutrinos. Besides that, extreme energy protons and ions yield
neutrinos through the GZK effect. 

\bigskip

All this information together should help to pinpoint the sources of
EECR. Whether the EECR are isotropic or not and whether their incoming
directions are correlated will be obviously a crucial information.

It would be then possible to decide whether the sources of EECR
are big objects like galaxies and their AGN, medium size objects like
the fireballs producing gamma ray bursts (GRB) or small objects like magnetars
(see discussion above). 

\section{Where to place the theoretical effort ?}

We will place our effort in the Bottom - up context, for instance:

\begin{itemize}

\item{1)} to extend the theory of Diffusive Shock Acceleration (DSA), 
(which  explains well the power law spectrum of standard CR's  over at 
least thirteen orders of magnitude), in order to incorporate:

Non-linear phenomena still poorly understood, i.e. back reaction of 
the particle on the plasma (on the shock wave), turbulence effects,
careful derivation of the transport equations and the coefficients in them,
resonant Alfven waves and 
dynamical instabilities; Confinement problem of the particles in the 
shock; gain of energy (minimal losses)

\item{2)}  Astrophysical Sources of EECR.

Our  aim is  to clarify the situation  on the currently proposed
models for EECR's, (`bottom-up' and `top-down' scenarii). The diffusive
shock  acceleration mechanisms allow to accelerate charged particles
like protons and ions, till energies only restricted by the size and the
magnetic field strength of the source. Active Galactic Nuclei, BL
Lacs, fireballs associated to GRB's, neutron stars appear as major
candidates for EECR. Our task is 
to improve the research on the potential astrophysical sources,
and to develop the theoretical research on the acceleration
mechanisms involving plasma physics, magnetohydrodynamics,
astrophysical shock waves and nonlinear phenomena still poorly understood.
A crucial task will be to single out features of the EECR and
prediction of its spectrum in order to discriminate among the
different possible sources. Mainly to distinguish among
`small' sources with high magnetic field (magnetars-young neutron
stars) and `medium' size sources (fireballs producing GRB's) from
`large' sources as radio galaxies and AGN.
\end{itemize}

\bigskip

In summary, the standard model of cosmic ray acceleration (diffusive
shock acceleration) based on Fermi ideas explains the non-thermal
power energy spectrum of CR over at least thirteen orders of
magnitude. It is {\bf reasonable} to extend such spectrum to EECR and
this seems {\bf plausible}. However, stimulating physical and
astronomical problems remain to understand  and explain the CR
spectrum well below extreme energies. 

\bigskip

Acknowledgments:

We thank R. Dick and E. Waxman for useful correspondence.

\end{document}